# Deep Learning-Based Automatic Modulation Classification Using GRU Networks

Osman Tokluoglu[1*], Emin Keresteci[2]
[1]Department of Electrical and Electronics Engineering, AYBU, Ankara, Türkiye
[2]Department of Computer Engineering, TOBB ETU, Ankara, Türkiye
***Contact:** otokluoglu@aybu.edu.tr, phone +90-312-906-2217

***Abstract***— Automatic modulation classification (AMC) plays a critical role in modern wireless communication systems, particularly in non-cooperative scenarios where prior knowledge of the transmitted signal is unavailable. In this study, a gated recurrent unit (GRU)-based deep learning framework is investigated for the classification of digital modulation schemes by exploiting the temporal characteristics of received signals. The proposed approach operates directly on in-phase and quadrature (I/Q) signal representations and aims to learn discriminative features in a data-driven manner without relying on handcrafted feature extraction. The performance of the proposed model is evaluated for BPSK, QPSK, and 16PSK modulation schemes under additive white Gaussian noise (AWGN) channel conditions across a wide range of signal-to-noise ratio (SNR) levels. The obtained results demonstrate that the GRU-based model achieves reliable classification performance, with overall accuracy improving from 55.3% at −10 dB SNR to 98.5% at 15 dB SNR. In particular, the model exhibits strong performance at moderate and high SNR levels, while maintaining reasonable accuracy even under challenging low SNR conditions. These findings suggest that GRU-based architectures provide a promising and computationally efficient solution for modulation classification tasks. The presented results represent an initial step toward more comprehensive studies, including extensions to fading channel environments, additional modulation schemes, and real-time implementations using hardware platforms.

## I. INTRODUCTION

Automatic modulation classification (AMC) is a fundamental component in modern wireless communication systems, enabling adaptive transmission, spectrum monitoring, cognitive radio, and signal intelligence applications [1]. In scenarios where prior information about the transmitted signal is unavailable, accurate identification of the modulation scheme becomes essential for reliable demodulation and efficient spectrum utilization. Traditional AMC techniques can be broadly categorized into likelihood-based and feature-based approaches. Likelihood-based methods, such as maximum likelihood (ML) classifiers, offer optimal performance under ideal conditions but suffer from high computational complexity and sensitivity to model mismatches [2], [3]. On the other hand, feature-based methods rely on expert-designed statistical features, including higher-order cumulants and cyclostationary properties, which reduce computational burden but often exhibit degraded performance in low signal-to-noise ratio (SNR) environments and under channel impairments [4], [5].

In recent years, the rapid advancement of deep learning has led to a paradigm shift in AMC, where data-driven models automatically learn discriminative features directly from raw or minimally processed signals. Early works demonstrated the effectiveness of convolutional neural networks (CNNs) in extracting spatial features from in-phase and quadrature (I/Q) samples [6]. Subsequent studies further improved classification performance by employing deeper CNN architectures and hybrid models combining convolutional and fully connected layers [7]. However, while CNN-based approaches are effective at capturing local patterns, they are inherently limited in modeling long-range temporal dependencies present in communication signals [8].

To address this limitation, recurrent neural networks (RNNs) have been introduced for AMC tasks. These models are designed to capture sequential dependencies in time-series data, making them well-suited for communication signals where temporal correlations carry important modulation-specific information. For instance, a long short-term memory (LSTM)-based AMC framework that extracts robust features from noisy radio signals to efficiently and accurately classify modulation schemes and communication technologies is proposed in [9]. In another study, a hybrid approach employing a dual-stream CNN–LSTM framework for automatic modulation classification that jointly exploits spatial–temporal features and feature interactions from I/Q and amplitude–phase representations to improve performance is proposed in [10].

In this context, gated recurrent unit (GRU)-based architectures offer a promising alternative for modulation classification due to their ability to model temporal dependencies with a relatively simpler structure compared to other recurrent models. Motivated by these advantages, this study investigates the use of a GRU-based framework for automatic modulation identification. By leveraging the sequential characteristics of communication signals, the proposed approach aims to effectively capture temporal patterns relevant to different modulation schemes while maintaining computational efficiency.

The remainder of this paper is organized as follows: Section II presents the system model and problem formulation. Section III describes the proposed GRU-based modulation identification framework. Section IV provides the simulation setup and discusses the obtained results. Finally, Section V concludes the paper and outlines potential directions for future work.

## II. System Model

In this study, the AMC problem is formulated as a supervised classification task, where the objective is to determine the modulation scheme of a received signal from a predefined set of candidates. Specifically, the considered modulation types are Binary Phase Shift Keying (BPSK), Quadrature Phase Shift Keying (QPSK), and 16-Phase Shift Keying (16PSK), which represent commonly used phase modulation techniques with increasing constellation complexity.

### A. Transmitted Signal Model

Let $\{a_k\}$ denote the sequence of transmitted symbols drawn from the corresponding modulation constellation. For BPSK, QPSK, and 16PSK, the symbols are defined as complex-valued points on the unit circle, expressed as $a_k = e^{j\theta_k}$, where $\theta_k$ represents the phase associated with the selected modulation scheme. Specifically, the phase sets are given by:

- BPSK: $\theta_k \in \{0, \pi\}$,
- QPSK: $\theta_k \in \left\{0, \frac{\pi}{2}, \pi, \frac{3\pi}{2}\right\}$,
- 16PSK: $\theta_k \in \left\{\frac{2\pi m}{16} \mid m = 0,1,\dots,15\right\}$.

The transmitted baseband signal can be written as

$$s(t) = \sum_k a_k g(t - kT), \tag{1}$$

where $g(t)$denotes the pulse shaping filter and $T$is the symbol duration.

### B. Channel Model

The transmitted signal propagates through a wireless channel and is impaired by additive noise. In this study, the channel is modeled as an additive white Gaussian noise (AWGN) channel, and the received signal is given by

$$r(t) = s(t) + n(t), \tag{2}$$

where $n(t)$represents complex-valued white Gaussian noise with zero mean and variance $\sigma^2$. After matched filtering and sampling at the symbol rate, the discrete-time received signal can be expressed as

$$r_k(t) = a_k(t) + w_k(t), \tag{3}$$

where $w_k \sim \mathcal{CN}(0, \sigma^2)$ denotes the discrete-time noise component.

### C. Signal Representation

For deep learning-based processing, the received signal is represented in terms of its in-phase (I) and quadrature (Q) components. Specifically, each received symbol is decomposed as

$$r_k(t) = I_k + jQ_k, \tag{4}$$

where $I_k = \Re\{r_k\}$and $Q_k = \Im\{r_k\}$. A sequence of $N$received samples is then organized into a two-dimensional input representation given by

$$\boldsymbol{X} = [I_1, Q_1, I_2, Q_2, \dots, I_N, Q_N], \tag{5}$$

or equivalently as an $N \times 2$matrix. This representation preserves the temporal structure of the signal and enables the model to learn modulation-specific patterns from the I/Q components.

### D. Problem Formulation

Given the input sequence **X**, the goal of the modulation classifier is to estimate the corresponding modulation label $y \in$ {BPSK,QPSK,16PSK}. This can be formulated as a multi-class classification problem, where a neural network model $f(\cdot)$is trained to learn the mapping $y = f(\boldsymbol{X})$. The model is trained using labeled data generated under different SNR conditions to ensure robustness against noise variations.

## III. Proposed GRU-Based Modulation Identification Framework

In this section, the proposed GRU-based modulation identification framework is presented. The overall objective is to exploit the temporal characteristics of the received signal to accurately classify different modulation schemes. The proposed model operates directly on the sequential in-phase and quadrature (I/Q) representation of the received signal and learns modulation-specific patterns in a data-driven manner.

### A. Input Representation

As described in Section II, the received signal is represented using its in-phase (I) and quadrature (Q) components. For each sample, a sequence of $N$symbols is considered and arranged as

$$\boldsymbol{X} = \{[I_1, Q_1], [I_2, Q_2], \dots, [I_N, Q_N]\}, \tag{6}$$

which forms an $N \times 2$input matrix. This representation preserves both the amplitude and phase information of the signal while maintaining its temporal structure. Such a formulation is particularly suitable for recurrent neural networks, which are specifically designed to process sequential data by capturing temporal dependencies and correlations between successive samples, thereby enabling the model to effectively learn the underlying structure and dynamic behavior of time-varying communication signals.

### B. GRU-Based Architecture

To model the temporal dependencies in the input signal, a neural network architecture based on GRUs is employed. Compared to conventional recurrent neural networks, GRUs incorporate gating mechanisms that help mitigate the vanishing gradient problem and enable more effective learning of long-range dependencies.

As illustrated in Fig. 1, the GRU unit consists of two main gates: the update gate and the reset gate [11]. These gates control the flow of information through the network and determine how much of the past information is retained or updated at each time step. Given an input sequence **X**, the GRU processes the data sequentially and produces a hidden representation that captures the temporal evolution of the signal.

In this study, the proposed architecture consists of one or more stacked GRU layers, followed by a fully connected layer for classification. The GRU layers extract temporal features from the I/Q sequence, while the dense layer maps these features to the corresponding modulation classes.

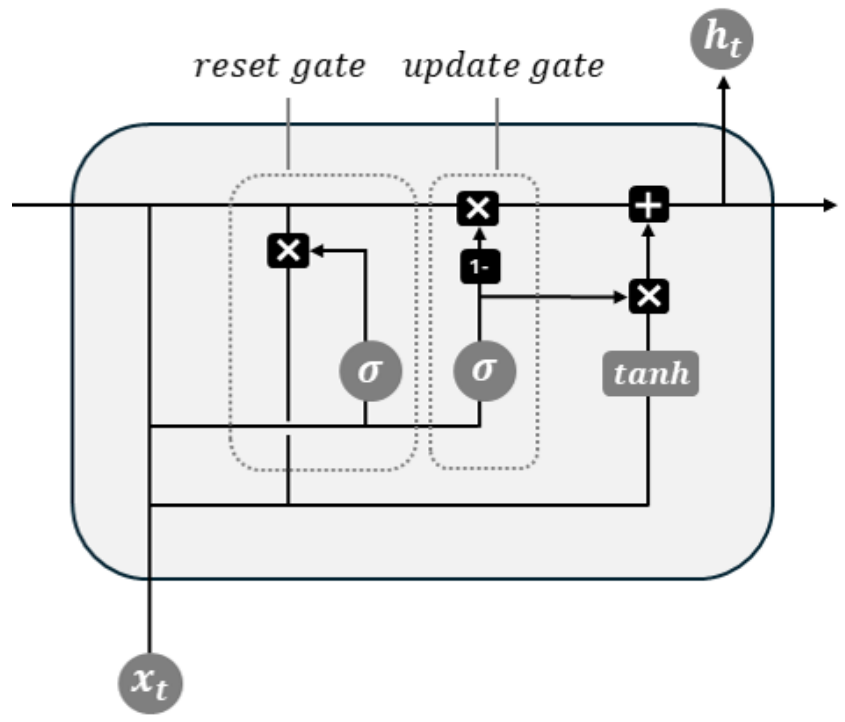

**Fig. 1** Structure of a gated recurrent unit (GRU).

### C. Classification Layer

The output of the final GRU layer is passed to a fully connected (dense) layer, which performs the classification task. A softmax activation function is used at the output layer to produce the probability distribution over the modulation classes, namely BPSK, QPSK, and 16PSK.

Let **h** denote the feature vector obtained from the GRU layers. The output of the classifier is given by

$$\hat{\boldsymbol{y}} = softmax(\boldsymbol{Wh} + \boldsymbol{b}), \tag{7}$$

where **W** and **b** represent the weights and bias of the dense layer, respectively, and $\hat{\mathbf{y}}$ denotes the predicted probability vector. The final decision is made by selecting the class with the highest probability, as determined by the output of the softmax layer, which assigns a likelihood value to each possible modulation class based on the learned feature representation.

### D. Training Procedure

The model is trained in a supervised manner using labeled data generated according to the system model described in Section II. The training dataset includes signals corresponding to different modulation schemes under various SNR conditions to improve the robustness of the model.

The categorical cross-entropy loss function is employed to measure the discrepancy between the predicted and true labels. The network parameters are optimized using a gradient-based optimization algorithm such as Adam.

To enhance generalization performance, the dataset is divided into training and validation sets. The training process is carried out over multiple epochs until convergence is achieved.

## IV. Simulation Results

In this section, the performance of the proposed GRU-based modulation identification framework is evaluated. The results are presented in terms of classification accuracy under different SNR conditions.

### A. Simulation Setup

The dataset is generated according to the system model described in Section II. The considered modulation schemes include BPSK, QPSK, and 16PSK. For each modulation type, a large number of signal sequences are generated with random symbol realizations.

The transmitted signals are passed through an additive white Gaussian noise (AWGN) channel, and different SNR levels are considered to evaluate the robustness of the model. In this study, the SNR range is selected from $\mathrm{SNR_{min}} = -10$ to $\mathrm{SNR_{max}} = 15$ dB with a step size of ΔSNR=5 dB.

Each received signal is represented using its in-phase (I) and quadrature (Q) components and organized into sequences of length $N$. The dataset is divided into training, validation, and test sets. The model is trained using the Adam optimizer with categorical cross-entropy loss.

### B. Performance Metric

The performance of the proposed model is evaluated using classification accuracy, defined as

$$\text{Accuracy} = \frac{\text{Number of correctly classified samples}}{\text{Total number of samples}}. \tag{8}$$

Accuracy is computed for each SNR level to analyze the behavior of the model under different noise conditions, allowing for a detailed evaluation of its robustness and classification performance as the signal quality varies from highly noisy to relatively clean scenarios, thereby providing a clearer understanding of its performance trends across different SNR regimes.

### C. Results and Discussion

The classification accuracy of the proposed GRU-based model under different SNR conditions is presented in Table I.

**TABLE I**
CLASSIFICATION ACCURACY (%) OF THE PROPOSED GRU-BASED MODEL AT DIFFERENT SNR LEVELS

| SNR (dB) | BPSK (%) | QPSK (%) | 16PSK (%) | Overall Accuracy |
|---|---|---|---|---|
| -10 | 68.5 | 55.2 | 42.3 | 55.3 |
| -5 | 82.7 | 71.4 | 58.6 | 70.9 |
| 0 | 92.8 | 85.3 | 73.9 | 84.0 |
| 5 | 97.6 | 93.8 | 86.5 | 92.6 |
| 10 | 99.1 | 97.5 | 92.8 | 96.5 |
| 15 | 99.8 | 99.1 | 96.7 | 98.5 |

As shown in Table I, the classification accuracy improves as the SNR increases, which is consistent with expectations since higher SNR values correspond to less noisy signal conditions and a clearer representation of the underlying modulation patterns. At low SNR levels, the performance degrades noticeably due to the increased difficulty in distinguishing modulation-specific features in the presence of strong noise. In such conditions, the noise components can significantly distort the amplitude and phase information of the received signal, leading to overlapping feature distributions among different modulation classes and consequently reducing classification accuracy.

Among the considered modulation schemes, BPSK generally achieves higher classification accuracy compared to QPSK and 16PSK, particularly at low SNR levels. This behavior can be attributed to the simpler constellation structure of BPSK, which consists of only two distinct symbols with maximum separation in the signal space, making it inherently more robust to noise and distortion. QPSK, while still relatively robust, exhibits slightly lower performance due to its increased constellation density. In contrast, 16PSK demonstrates comparatively lower accuracy, especially in low and moderate SNR regimes, due to its densely packed constellation points. The smaller angular separation between symbols in 16PSK makes it more susceptible to noise and phase variations, thereby increasing the likelihood of misclassification.

As the SNR increases, the performance gap between the modulation schemes gradually decreases, and all classes achieve high classification accuracy at moderate to high SNR levels. This indicates that the model is able to effectively learn and exploit distinguishing features when the signal quality is sufficiently high. Overall, the proposed GRU-based model demonstrates reliable classification performance across a wide range of SNR conditions. The results suggest that the model effectively captures temporal dependencies and modulation-specific patterns, highlighting the potential of GRU-based architectures for modulation classification in practical communication systems.

## V. CONCLUSIONS

In this paper, a GRU-based framework for automatic modulation identification has been investigated, focusing on the exploitation of temporal characteristics of communication signals. The obtained results demonstrate that the proposed approach is capable of achieving reliable classification performance for BPSK, QPSK, and 16PSK modulation schemes under AWGN channel conditions. These findings can be considered as an initial step toward understanding the potential of GRU-based models in modulation classification tasks. Future work will focus on extending the proposed framework to more realistic communication scenarios. In particular, the performance of the model under fading channel conditions will be investigated, along with its applicability to a wider range of modulation schemes. Additionally, practical implementations using software-defined radio (SDR) platforms, such as Universal Software Radio Peripheral devices, as well as hardware-oriented solutions including FPGA-based realizations, will be considered to evaluate the feasibility of the approach in real-world environments, with particular attention to real-time processing capabilities, computational efficiency, and deployment constraints in practical communication systems.

## REFERENCES

[1] O. A. Dobre, A. Abdi, Y. Bar-Ness, and W. Su, "Survey of automatic modulation classification techniques: Classical approaches and new trends," *IET Commun.*, vol. 1, no. 2, pp. 137–156, Apr. 2007.

[2] W. Wei and J. M. Mendel, "Maximum-likelihood classification for digital amplitude-phase modulations," *IEEE Trans. Commun.*, vol. 48, no. 2, pp. 189–193, Feb. 2000.

[3] J. L. Xu, W. Su, and M. Zhou, "Likelihood-ratio approaches to automatic modulation classification," *IEEE Trans. Syst., Man, Cybern. C, Appl. Rev.*, vol. 41, no. 4, pp. 455–469, Jul. 2011.

[4] A. Swami and B. M. Sadler, "Hierarchical digital modulation classification using cumulants," *IEEE Trans. Commun.*, vol. 48, no. 3, pp. 416–429, Mar. 2000.

[5] F. -X. Socheleau, "Cyclostationarity of communication signals in underwater acoustic channels," *IEEE J. Ocean. Eng.*, vol. 50, no. 2, pp. 425–447, Apr. 2025.

[6] T. J. O'Shea, J. Corgan, and T. C. Clancy, "Convolutional radio modulation recognition networks," arXiv:1602.04105, Feb. 2016.

[7] S. Rajendran, W. Meert, D. Giustiniano, V. Lenders, and S. Pollin, "Deep learning models for wireless signal classification with distributed low-cost spectrum sensors," *IEEE Trans. Cogn. Commun. Netw.*, vol. 4, no. 3, pp. 433–445, Sept. 2018.

[8] O. Tokluoglu, E. Cavus, E. Bedeer, and H. Yanikomeroglu, "A novel CNN-based standalone detector for faster-than-Nyquist signaling," *IEEE Trans. Commun.*, vol. 73, no. 12, pp. 14316–14331, Dec. 2025.

[9] Z. Ke and H. Vikalo, "Real-time radio technology and modulation classification via an LSTM auto-encoder," *IEEE Trans. Wireless Commun.*, vol. 21, no. 1, pp. 370–382, Jan. 2022.

[10] Z. Zhang, H. Luo, C. Wang, C. Gan, and Y. Xiang, "Automatic modulation classification using CNN-LSTM based dual-stream structure," *IEEE Trans. Veh. Technol.*, vol. 69, no. 11, pp. 13521–13531, Nov. 2020.

[11] O. Tokluoglu, A. Cicek, E. Cavus, E. Bedeer, and H. Yanikomeroglu, "GRU-based sequence detection for faster-than-Nyquist signaling," *IEEE Open J. Veh. Technol.*, vol. 7, pp. 565–581, 2026